\documentclass[11pt]{article}

\usepackage[utf8]{inputenc}
\usepackage[T1]{fontenc}
\usepackage[margin=1in]{geometry}
\usepackage{amsmath,amssymb}
\usepackage{booktabs}
\usepackage{graphicx}
\graphicspath{{figs/}}
\usepackage{siunitx}
\usepackage[hidelinks]{hyperref}
\usepackage{xcolor}

\newcommand{\detail}[1]{\smallskip\noindent\emph{Detail (skippable).} #1\smallskip}

\title{\textbf{Billion-Scale Nearest-Neighbor Search under Fully Homomorphic\\
Encryption on a Single GPU, Balancing Leakage and Cost}}
\author{
  Isamu Isozaki\thanks{\texttt{imi25@drexel.edu}} \\ \emph{Drexel University}
  \and
  Madison Bratina \\ \emph{Miriel AI}
  \and
  Edward Kim\thanks{\texttt{ek826@drexel.edu}} \\ \emph{Drexel University}
}
\date{\today}

\begin{document}
\maketitle

% =====================================================================
\begin{abstract}
We build a system that answers ``which database vectors are most similar to my
query?'' \emph{without the server ever seeing the query}. The query is encrypted
with fully homomorphic encryption (FHE); the server does all its scoring on
ciphertexts and returns encrypted results that only the client can read. The
challenge is speed: at a billion vectors, scoring every row under encryption is far
too slow, so we combine two ideas---\emph{rank reduction} (shrink each vector's
dimension) and a \emph{hierarchy} (route to a small candidate set instead of scanning
everything)---executed under encryption on a single GPU. We evaluate on three corpora at
very different scales: a face corpus of \num{222049} centroids clustered from
${\sim}10$\,M face images (512-dim), DataComp-1B ($1.39\times10^9$
vectors, 512-dim CLIP), and Deep1B ($10^{9}$ vectors, 96-dim). On DataComp-1B we reach a recall@10 of \textbf{0.90} against
the single labeled answer, or \textbf{0.95} when a near-duplicate image in the top-10
also counts as correct (the data is web-scraped and full of duplicates), at ${\sim}6$\,s
per encrypted query on a GPU; a lighter configuration reaches $0.78$/$0.83$ at
${\sim}1.8$\,s. These are warm (deployable) server-side latencies---client decryption and
network transfer are excluded. On Deep1B we reach recall@10 \textbf{0.90} under all-levels FHE
($0.9045$ measured over $2000$ FHE queries, matching the $0.906$ plaintext routing---the $96{\to}128$
zero-pad is exact, correlation $1.0$) at $2.3$\,s warm per query. We describe the full client--server protocol in enough detail to reproduce it,
and report accuracy and latency for every configuration. We also measure what this speed
costs: the hierarchy's access pattern leaks the database geometry (an observer recovers
$72\%$ of the coarse-cell neighbor graph from access patterns alone), and we show that
seeded (fixed-group) padding cuts this leak by ${\sim}35\times$ (to ${\approx}2\%$), where
naive padding is defeated by a repeated-query attack.
\end{abstract}

% =====================================================================
\section{Introduction}
\label{sec:intro}
Suppose you want to search a large database of vectors---image embeddings, face
templates---but the query is private and the database lives on an untrusted server.
Fully homomorphic encryption (FHE) lets the server compute on encrypted data, so it
can score your encrypted query against its database and hand back an encrypted answer
it cannot read. The catch is cost: FHE arithmetic is expensive, and a database can
have a billion entries. Scanning all of them under encryption per query is
impractical: a prior GPU system brute-force-searched $10$\,M encrypted face vectors on
\emph{eight} GPUs~\cite{miriel10m}, so extrapolating linearly to DataComp's $1.39$\,B
vectors would take on the order of $1100$ GPUs---whereas the hierarchical approach here
runs on a \emph{single} GPU. The real question is thus how to score only a
\emph{small, well-chosen} candidate set---under encryption---while keeping accuracy high.

This paper is a systems paper: we show how to make encrypted approximate
nearest-neighbor (ANN) search work at billion scale on a GPU, and we report what it
costs and how well it works. Our contributions:
\begin{itemize}
  \item A \textbf{clear, reproducible end-to-end protocol} for encrypted ANN
        (\S\ref{sec:overview}): what the client sends, what the server computes on
        ciphertexts, and what comes back.
  \item \textbf{Rank reduction under encryption} (\S\ref{sec:method:rank}): the
        server projects the encrypted query to a smaller dimension without ever
        seeing the projection matrix. We also solve a subtlety this creates---the
        projection distorts vector norms, so recovering true cosine similarity takes
        care---and prove a packing optimization correct.
  \item An \textbf{all-levels-FHE hierarchical search} (\S\ref{sec:method:hier}):
        coarse\,$\to$\,fine\,$\to$\,rerank routing where each level is scored under
        encryption and the client steers, so a query touches ${\sim}65$--$110$k candidates
        instead of $1.39$\,B.
  \item A set of \textbf{GPU optimizations} (\S\ref{sec:method:gpu}) that cut the
        warm per-query cost by ${\sim}5.4\times$.
  \item A \textbf{full evaluation} (\S\ref{sec:eval}): accuracy and latency for every
        configuration on all three corpora.
  \item An \textbf{access-pattern leakage analysis} (\S\ref{sec:leakage}): plaintext
        metrics for how much the hierarchy's co-access reveals of the database geometry
        and the query stream, and how padding---undone by a repeated-query intersection
        attack unless the decoys are fixed---trades leak for fetch cost.
\end{itemize}
We also explored product quantization under FHE but do \emph{not} use it in our DataComp
results; \S\ref{sec:method:pq} explains why (a real negative finding: fully hidden it does
not beat plain encrypted search, and the leakage-allowing variant leaks too much).

% =====================================================================
\section{System Overview: one query, start to finish}
\label{sec:overview}
This section is the map for the rest of the paper. Read it and you know how the
system works.

\paragraph{Setting.} This overview describes the billion-scale (DataComp and Deep1B) path, which is
where the hierarchy applies; the smaller face corpus is scored flat (no hierarchy,
\S\ref{sec:eval:face}). The client owns the data and a CKKS secret key. Offline, the
data owner clusters the database into a \emph{hierarchy}---coarse cells, then finer
cells, then leaves holding the actual rows---and gives the server the (encrypted)
cluster centroids and the (encrypted) projection matrix. At query time the client
holds the secret key; the server holds only encrypted things and does encrypted
arithmetic (in a deployment---our prototype encrypts on the fly; see below). This is the
standard single-key, outsourced-database setting (the querier is the data owner, not a
third party).

\paragraph{Measured prototype vs.\ deployment.} We state this plainly because it affects
the threat model. Our \emph{measured} system encrypts the touched database rows
\emph{on the fly} at query time---this is the per-query ``enroll (encrypt)'' term in the
DataComp latency breakdown (Fig.~\ref{fig:datacomp})---so during our runs the database
sits in \emph{plaintext} on the server and is encrypted per query. That is a prototype
convenience, not the setting above. We therefore report two latencies: a \emph{cold}
number that includes this on-the-fly per-query encryption (the prototype cost), and the
deployable \emph{warm} number---measured after the candidate rows are enrolled and their
ciphertexts serialized to disk, so it is the cost of loading those pre-encrypted
ciphertexts back from disk and scoring them---which is what Table~\ref{tab:datacomp} quotes; the cold number is roughly $3.5\times$ larger (e.g.\
$13.8$ vs.\ $4.0$\,s at R256/leaf-$32$). A deployment matching
the threat model pre-encrypts the database once, offline, and stores ciphertexts; there
is then no per-query enroll cost, only a one-time storage cost, and it follows directly from the CKKS parameters
($128$-bit security, $45$-bit scale). A serialized ciphertext is
$2\,{\cdot}\,L\,{\cdot}\,n\,{\cdot}\,8\,{\cdot}\,2.5$ bytes ($L$ RNS limbs, ring dimension
$n$, $8$ bytes per coefficient, ${\sim}2.5\times$ serialization overhead); it holds $n/2$
slots, and a block of $n/2$ vectors is stored as $R$ diagonals (one per reduced
dimension), so the per-vector cost is
\[
  \frac{R\,\cdot\,2 L n\,\cdot\,8\,\cdot\,2.5}{n/2}\;=\;80\,L\,R \ \text{bytes},
\]
independent of the ring dimension $n$. At the shallow level the ranked matvec needs
($L{=}5$, i.e.\ level $4$) this is $80\,{\cdot}\,5\,{\cdot}\,128\approx 51$\,KB per vector
at $R{=}128$ ($\approx 102$\,KB at $R{=}256$). Hence the ${\sim}10^{6}$-centroid routing
index is ${\approx}51$\,GB, and the full $1.39$\,B rows are
$1.39{\times}10^{9}\,{\cdot}\,51\,\text{KB}\approx 71$\,TB ($\approx 142$\,TB at $R{=}256$).
A deployment stores all of this as ciphertext---both the routing index and the reranked
base rows are pre-encrypted at rest---so the $71$\,TB is a one-time storage cost; our
prototype encrypts on demand only to avoid paying it during the measured runs.
The ranked path needs \emph{no} homomorphic thresholding (the client ranks after
decryption), so the database can live at this low level; the membership (face) path
instead evaluates an encrypted threshold (${\sim}11$ levels, $L{\approx}12$), which by the
same formula stores ${\sim}2.4\times$ ($12/5$) larger per vector.

\paragraph{Security parameters.} Both parameter sets are $128$-bit classic (OpenFHE
\texttt{HEStd\_128\_classic}, $45$-bit scale), with the library choosing the ring
dimension: the ranked DataComp path runs at multiplicative depth $4$ (ring $2^{14}$) and
the face membership path at depth $11$ (ring $2^{15}$, which the deeper chain needs to
stay $128$-bit).

\paragraph{What CKKS gives us.} CKKS is an FHE scheme for approximate real-number
arithmetic. You can add and multiply ciphertexts, and---because it packs many numbers
into one ciphertext (``slots'')---rotate the packed vector. Multiplications are the
scarce resource: each ciphertext has a small \emph{multiplicative depth} budget, so
the whole pipeline is designed to spend as few multiplications as possible.

\begin{figure}[t]
  \centering
  \includegraphics[width=0.92\linewidth]{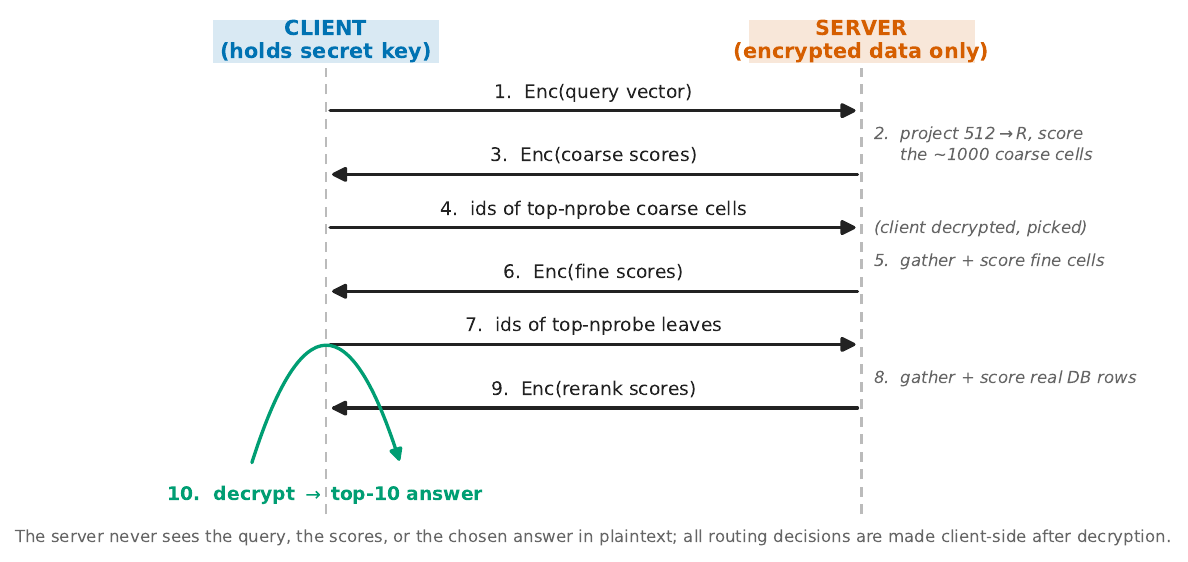}
  \caption{One encrypted query, start to finish. The client (left) holds the secret
  key and makes all routing decisions; the server (right) only stores encrypted data
  and computes on ciphertexts.}
  \label{fig:pipeline}
\end{figure}

\paragraph{The round trip.} Figure~\ref{fig:pipeline} shows one query end to end:
\begin{enumerate}
  \item \textbf{Client sends} the query vector, encrypted (the server never sees it
        in the clear).
  \item \textbf{Server, coarse level:} it homomorphically projects the encrypted
        query to a smaller rank $R$ and scores it against the ${\sim}1000$ coarse
        centroids (an encrypted matrix--vector product), returning the encrypted
        scores.
  \item \textbf{Client:} decrypts the coarse scores, keeps the top few cells
        (``\texttt{nprobe}''), and tells the server which ones.
  \item \textbf{Server, fine level:} scores the query against the finer centroids
        under those cells; returns encrypted scores.
  \item \textbf{Client:} decrypts, keeps the top leaves.
  \item \textbf{Server, rerank:} gathers the \emph{actual} database rows in those
        leaves, scores them, returns encrypted scores.
  \item \textbf{Client receives:} decrypts, takes the top-10 $\to$ the answer.
\end{enumerate}
The server only ever does encrypted arithmetic; every ``which cells next'' decision
happens on the client after it decrypts. This is why the server is a simple,
stateless scorer and all the routing logic lives in the client.

\paragraph{Reproducing this.} The server is a persistent process that loads the CKKS
keys once and then, per request, receives a level id plus a list of group ids and the
encrypted query, gathers those rows, scores them under encryption, and returns the
score vector. Concretely we use ring dimension $16384$, a shallow multiplicative
depth (4), and score in batches of $8192$ rows; the client packs candidate ids into
$\le\!8192$-row batches. The two levers that actually make the per-request scoring cheap are
rank reduction and the hierarchy; we also study product quantization (PQ), but---as a key
negative result---find it does \emph{not} pay under encryption (\S\ref{sec:method}).

% =====================================================================
\section{Method}
\label{sec:method}
One of our techniques is \emph{compression} (make each vector cheaper: rank
reduction) and one is \emph{structure} (score fewer vectors: the hierarchy); we also
analyze product quantization---a second, natural compression---but find it does
\emph{not} pay under FHE, and do not use it. We take them in turn, each time giving
the plain idea first and the FHE mechanics second.

\subsection{Rank reduction: project to fewer dimensions}
\label{sec:method:rank}
\paragraph{Idea.} A $512$-dim vector is expensive to score under encryption. If we
first project it to $R$ dimensions ($256$ or $128$), scoring gets cheaper, and---as
long as $R$ isn't too small---accuracy barely drops. The twist is that the projection
must happen \emph{under encryption} (the query is encrypted) and \emph{without the
server learning the projection matrix} $W$.

\paragraph{How.} The server holds $W$ as encrypted ``diagonals'' and computes
$\mathrm{Enc}(q_R)=W^{\!\top}\mathrm{Enc}(q)$ with the standard diagonal
matrix--vector method (encode a matrix by its diagonals; a matrix--vector product
becomes a few rotations and multiplies). We pick the projection basis by Fisher
discriminant analysis over the clusters (``lda-clusters''), which separates classes
better than plain PCA/SVD for high-dimensional data.

\paragraph{The norm problem.} Projection does not preserve length: a unit vector,
projected, comes out with some other magnitude. So the raw encrypted score is
$|q_R|\cdot\cos(\theta)$, not the cosine we want. The database vectors can be
re-normalized cheaply \emph{in plaintext} when they are enrolled, but the query
cannot---after projection it exists only as a ciphertext, and dividing an encrypted
vector by its own (encrypted) length is expensive. How much this matters depends on
what the server returns:
\begin{itemize}
  \item \textbf{Ranked retrieval (our hierarchy).} The server returns the raw scores
        and the client picks the top-$k$. The client owns its query and the
        projection, so it simply computes $|q_R|$ in plaintext and divides after
        decrypting---exact, and essentially free. This is the path our DataComp
        results use.
  \item \textbf{Membership search (the face task).} Here the server must return a
        yes/no match \emph{under encryption}---there is no decrypt-then-normalize
        step. We instead compare against a threshold scaled into the query's units:
        approximate $\tau\,|q_R|=\sqrt{\tau^2|q_R|^2}$ with a \emph{low-degree}
        Chebyshev polynomial (the square root is smooth, so a low degree suffices, and
        we use a direct evaluator because the standard Paterson--Stockmeyer method is
        unstable at these tiny degrees), then test
        $\mathrm{sign}(\text{score}-\tau|q_R|)$. Since multiplications are scarce we
        pre-scale into $[-1,1]$ with a plaintext constant, which also saves one
        multiplication. This threshold path runs at multiplicative depth $11$, matching
        the prior GPU encrypted-matching work~\cite{demicheli2026encface}---the
        sign/threshold approximation is the deep part---except that the norm-recovery
        and the sign check now \emph{share} that budget, so we run the Chebyshev at a
        \emph{lower degree} than that work.
\end{itemize}
By contrast the ranked hierarchical path returns raw scores and needs no homomorphic
threshold, so it runs shallow (multiplicative depth $4$)---the difference between the
two paths is exactly whether the client or the server does the final normalize/threshold.

\begin{figure}[t]
  \centering
  \includegraphics[width=0.55\linewidth]{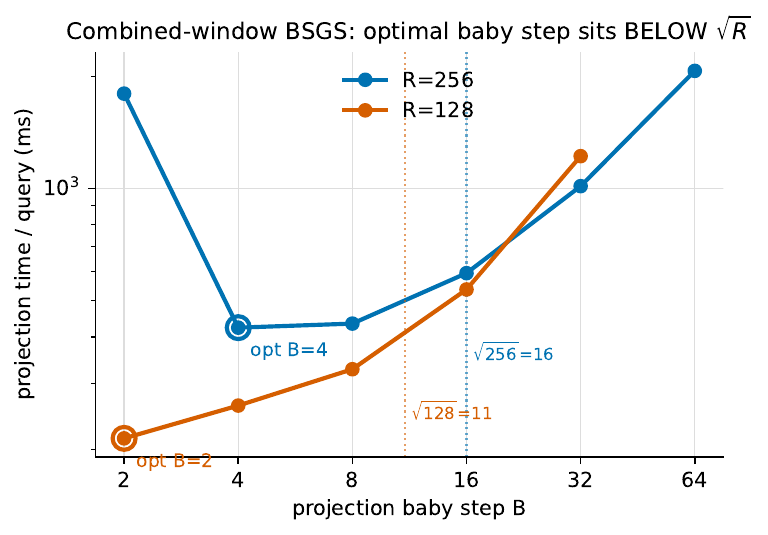}
  \caption{Projection latency vs.\ baby step $B$. The optimum is below $\sqrt{R}$
  because our packing makes the baby rotations recur per chunk while the giant
  rotation is shared.}
  \label{fig:babystep}
\end{figure}

\detail{A packing optimization and its correctness. We pack all chunks of $W$ into
one ciphertext using non-overlapping windows, cutting projection storage by a factor
$c=512/R$ at no extra homomorphic cost, and combine chunks with cheap \emph{plaintext}
$0/1$ masks (public constants, so applying them is a cheap ciphertext--plaintext
multiply with no key-switch, and masking never moves data across window boundaries
the way a rotation would). Under baby-step/giant-step (BSGS) the giant rotation acts
on the already-combined ciphertext and mixes adjacent windows; nonetheless a single
final fold recovers the exact answer (the mixing is a fixed permutation, invisible to
the final sum), whereas folding after each giant step is provably wrong. A knock-on
effect: because in this packing the baby rotations recur per chunk while the giant
rotation is shared, the best baby step sits \emph{below} the textbook $\sqrt{R}$
(measured optimum $B{=}4$ at $R{=}256$, $B{=}2$ at $R{=}128$; Fig.~\ref{fig:babystep}).
One more enrollment trick matters for the hierarchy, where a fresh candidate set is
encrypted every request: we \emph{plaintext-prerotate} the database---shift each database
diagonal by the giant-step amount \emph{in plaintext} at enrollment---so the server needs
no database rotation keys at query time (a rotation-free load, used in all reported latencies). This removes a per-request key-switch and fixes a crash where re-encrypting
under a freshly generated key would otherwise free the rotation keys the next request
needs.}

\subsection{Hierarchy: score a few thousand rows, not a billion}
\label{sec:method:hier}
The hierarchy is used \emph{only} for the billion-scale corpora (DataComp and Deep1B); the $222$k-centroid
face corpus is small enough to score exhaustively (a flat membership scan, \S\ref{sec:eval:face}),
so it uses no hierarchy, no routing, and no access-pattern leakage.
\paragraph{Idea.} Instead of scoring all $1.39$\,B rows, cluster them into a tree and
only descend into the promising branches. As in \S\ref{sec:overview}, the client
scores $\sim\!1000$ coarse cells, opens the best few, scores their fine cells, opens
the best leaves, and finally reranks the real rows in those leaves. A query touches
$\sim\!65$--$110$k rows total ($\sim\!1000$ coarse $+\;\sim\!18$k fine $+\;45$--$90$k
rerank), never the full billion.

\begin{figure}[t]
  \centering
  \includegraphics[width=0.55\linewidth]{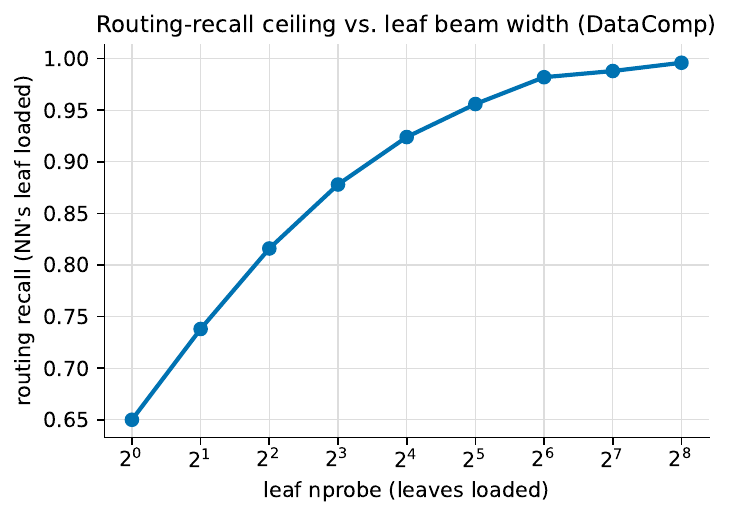}
  \caption{How often the true nearest neighbor's leaf is among the loaded leaves, as
  the leaf beam widens---the ceiling on rerank recall set by routing.}
  \label{fig:leaf-recall}
\end{figure}

\paragraph{How.} Each level is one call to the same stateless encrypted scorer: the
client sends the level's candidate ids and the encrypted query; the server gathers
those rows from a single per-level file (by offset and length), scores them, and
returns the encrypted scores. The client decrypts and chooses what to open next. Two
dials matter: the coarse beam width (which saturates---opening more than $16$ coarse
cells stops helping) and the leaf beam width, the main recall dial
(Fig.~\ref{fig:leaf-recall} shows how often the true neighbor's leaf is loaded as the
leaf beam widens).

\paragraph{The DataComp index, concretely.} For DataComp-1B the hierarchy is
\emph{two} levels, and it is worth stating exactly how it is built and queried---the
querying is not a deep balanced-tree descent to single-vector cells. \emph{Build:} we
cluster the $1.39$\,B base vectors into $10^{6}$ leaf clusters with a flat $k$-means;
each leaf is one cluster of base rows, represented by its centroid, so ``the leaves''
\emph{are} the $10^{6}$ centroids. We then build a coarse level of ${\sim}1000$ cells
on top by assigning every leaf centroid to its nearest coarse centroid (the coarse
level exists only to avoid scoring all $10^{6}$ leaves on every query). \emph{Query}: a search is exactly two encrypted rounds---(i)
score the encrypted query against the $1000$ coarse centroids and keep the top
$\mathrm{nprobe}_c{=}16$ cells; (ii) score it against the ${\sim}18$k leaf centroids
that live under those $16$ cells and keep the top $\mathrm{nprobe}_\ell$ ($32$ or $64$); then load
the base rows of those leaves and rerank in the projected $R$-dim. We fix the coarse beam
at $16$ throughout (both the recall and the leakage experiments); the leaf beam ($32$ or
$64$) is the main recall dial. Concretely, in the \emph{plaintext} recall
sweep the second round is implemented by scoring all $10^{6}$ leaf centroids and masking every
one \emph{not} under a chosen coarse cell to $-\infty$ before the top-$\mathrm{nprobe}_\ell$ (a
vectorization convenience, identical in result to scoring only the ${\sim}18$k candidates); the
\emph{deployed} FHE system instead fetches and scores only those ${\sim}18$k leaves under the
chosen coarse cells---which is exactly the first reveal the leakage model analyzes
(\S\ref{sec:leakage}). This flat, two-level scheme---a
coarse index over a single flat leaf level---is what the FHE pipeline runs; the deeper
balanced routing tree (splitting all the way down to one centroid per cell) is a
separate construction we do \emph{not} use here.

\subsection{Product quantization: explored, but not used for DataComp}
\label{sec:method:pq}
PQ replaces each vector by a short code that indexes small learned codebooks, so scoring
becomes cheap table lookups. We explored it under FHE but ultimately \emph{do not} use it
in the DataComp results, and why is a finding in its own right.

\paragraph{Fully encrypted, PQ does not pay.} Under encryption, PQ scoring has two phases.
The first---the dot product between the query's chunk embeddings and the codes---is
constant-time and cheap. The second---the \emph{gather}, mapping each code to its codebook
entry---is the problem: done under encryption it needs an encrypted one-hot vector per code
per centroid, which blows up storage and adds a ciphertext--ciphertext multiply per
centroid. Residual (additive) PQ softens this: with $K{=}2$ codewords per level and $M{=}2$
subquantizers, a centroid is reconstructed as a sum over $L$ residual levels,
\[
  c_i \;\approx\; \bigoplus_{m=1}^{M}\ \sum_{l=1}^{L}\ \mathrm{CB}^{(m)}_{l}\!\left[k^{(m)}_{i,l}\right],
\]
where $\oplus$ concatenates the $M$ chunks and each $(l,m)$ selects one of $K{=}2$ codewords,
so the one-hot per chunk shrinks from a flat $K'{\approx}2000$ to just $L$ two-way selects.
But now there is a ciphertext--ciphertext multiply at \emph{every} level, and the gather
still scales \emph{linearly} in the number of centroids scored---so, fully hidden, PQ never
beat plain encrypted vector search in our measurements.

\paragraph{Allowing leakage leaks too much.} The alternative is to offload the gather to the
client: the server returns a small distance table (constant-time, size independent of the
database) and the client does the lookup-and-sum in plaintext, which is $O(n)$ but cheap
because the client can index its own codes. But this is, in the worst case, a setting where
the attacker holds the \emph{plaintext codes without the codebook} (and without the original
data), so we can measure exactly how much the codes leak: rank neighbors by Hamming distance
on the codes, compare to the true nearest neighbors by cosine on the real centroids, and
report the fraction of each centroid's true cosine top-$10$ recovered from code agreement
(\emph{db-leak}@$10$; the attacker holds the codes, not the codebook). On the $10^{6}$
DataComp leaf centroids, publishing the residual codes at a single level leaks
db-leak@$10\approx 0.27$; running the coarse $1000$-way level \emph{under encryption} and
publishing only the second-level residual codes drops this to $0.075$ (shared codebook).
This residual client-gather is genuinely fast---its server cost is table-lookup and
independent of $N$ (${\approx}1.2$\,s)---but it buys that speed by publishing a
neighbor-graph leak the exact hierarchy does not have. Since
the
hierarchical index is already fast and leaks only its \emph{access pattern} rather than the
whole neighbor graph (\S\ref{sec:leakage}), we saw no benefit to the PQ path for DataComp and
dropped it.

\subsection{GPU optimizations for the encrypted scorer}
\label{sec:method:gpu}
\begin{figure}[t]
  \centering
  \includegraphics[width=0.7\linewidth]{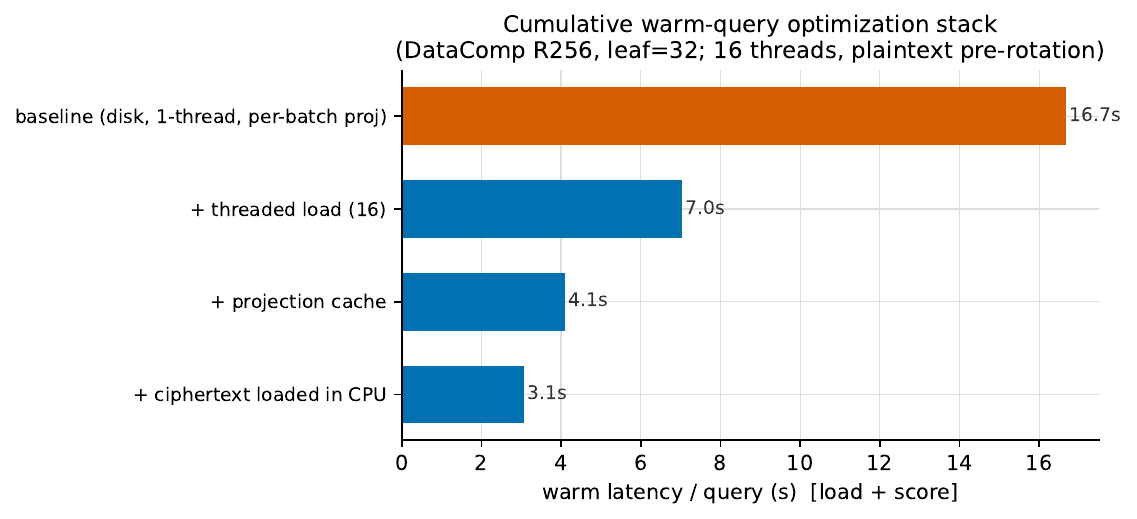}
  \caption{Cumulative effect of the software optimizations on the \emph{warm} per-query
  latency (load $+$ score, one-time enrollment excluded) of one DataComp query
  (R256, leaf=32; $16$ threads, plaintext pre-rotation throughout): from a disk,
  single-threaded, per-batch-projection baseline ($16.7$\,s) to the full stack
  ($3.1$\,s, in-mem), a $5.4\times$ speedup. The \emph{+\,projection cache} stage
  (disk load, plaintext pre-rotation) is exactly the Table~\ref{tab:datacomp}
  R256/leaf-$32$ configuration and matches its $4.0$\,s warm latency.}
  \label{fig:ablation}
\end{figure}

The encrypted matrix--vector product is the inner loop, and most of its cost turned
out to be GPU memory management, not arithmetic. The big wins, cumulatively cutting a
DataComp query's \emph{warm} latency (load $+$ score) from $16.7$ to $3.1$\,s
(Fig.~\ref{fig:ablation}): read the encrypted database with many threads; project the
query once and reuse it across batches; and hold the encrypted rows in CPU memory and
upload them straight to the GPU, instead of round-tripping them through disk.

\paragraph{Warm-query kernel path.} Two further GPU-kernel changes target the
\emph{warm} query---the steady-state latency once the one-time caches are hot, which is
the number that matters for a served system. First, we keep the small-step (baby-step)
rotation keys resident in VRAM: a cold query still generates them (${\sim}0.5$\,s), but
every subsequent warm query finds the key-init step essentially free. On its own,
though, caching exposed a second bottleneck---the many separate \texttt{malloc}s for
per-key ciphertexts fragmented GPU memory and slowed the following matrix multiply,
cancelling much of the gain. Allocating every ciphertext from one unified region
(\texttt{single\_alloc}), together with preallocating and reusing the temporary
ciphertexts, removes the fragmentation. Together these take the warm-query baby-step
path from $524$ to $22$\,ms (a ${\sim}96\%$ drop) and the full warm query from $813$ to
$323$\,ms on a brute-force scan of $2^{14}$ encrypted rows (Table~\ref{tab:warmopt}).
Beyond one ciphertext block---at $2^{15}$ rows, where a second set of $512$ diagonals is
needed---the BSGS scan itself dominates and the remaining gain is smaller ($1.65$ to
$1.20$\,s); the optimal baby step there was \texttt{BABY\_STEP\_N1}$=23$. These gains are
within noise on the DataComp hierarchy, whose per-query cost is dominated by enrollment
and rerank rather than a long warm scan, but they set the steady-state latency on the
membership (face) path.

\begin{table}[t]
  \centering\small
  \begin{tabular}{llrrr}
    \toprule
    Rows & Metric & Before & After & Reduction \\
    \midrule
    $2^{14}$ & warm query        & $813$\,ms  & $323$\,ms  & $60\%$ \\
    $2^{14}$ & \quad baby-step    & $524$\,ms  & $22$\,ms   & $96\%$ \\
    $2^{14}$ & total incl.\ setup & $9.68$\,s  & $5.60$\,s  & $42\%$ \\
    $2^{15}$ & warm query        & $1.65$\,s  & $1.20$\,s  & $27\%$ \\
    \bottomrule
  \end{tabular}
  \caption{Warm-query latency after small-step rotation-key VRAM caching and the
  \texttt{single\_alloc} fragmentation fix, on a face-scale brute-force encrypted scan.}
  \label{tab:warmopt}
\end{table}

% =====================================================================
\section{Evaluation}
\label{sec:eval}
We report accuracy and latency for every configuration on all three corpora, all measured on
a \emph{single} NVIDIA L40S (48\,GB) GPU on an AWS \texttt{g6e.4xlarge} instance
($16$ vCPUs = $8$ physical cores of an AMD EPYC 7R13, $128$\,GB host RAM, and an
${\sim}8$\,TB EBS SSD volume holding the pre-encrypted database, which the warm-query load streams
with $16$ threads at ${\sim}1$\,GB/s---a single stream reads EBS at only ${\sim}150$\,MB/s, so the
threading is what keeps the ciphertext load off the critical path). Every
FHE score was checked against the plaintext computation and matched to
${\sim}10^{-4}$, so the FHE and plaintext accuracies are identical---we therefore use
plaintext routing to sweep many configurations and FHE runs to measure latency. The
corpora stress different techniques, and their evaluation protocols differ
accordingly.

\paragraph{Datasets and queries.} All three corpora are evaluated on $2000$ held-out queries.
The \textbf{face corpus} is Glint360K (${\sim}10$\,M face images) clustered into
$\num{222049}$ centroids, each labeled by the majority face identity of its members; the
$2000$ queries are held-out face embeddings scored against all $\num{222049}$ centroids,
and a match is judged by identity (a query may match many same-identity centroids). For
\textbf{DataComp-1B} we build a disjoint random held-out split (seed $0$): $10{,}000$
image items are drawn and \emph{removed from the database} so a query cannot retrieve
itself; the reported recall and latency use $2000$ of these image-to-image queries (the
leakage study in \S\ref{sec:leakage} uses all $10{,}000$), and ground truth is each
query's true nearest database image. For \textbf{Deep1B} (the standard billion-scale ANN benchmark:
$10^{9}$ deep image descriptors, 96-dim, L2-normalized) we cluster the base into $10^{6}$ leaf
centroids under a $1000$-cell coarse level, and use $2000$ held-out queries from the public query set
with the official rank-1 nearest neighbor as ground truth; being $96$-dim, its vectors are zero-padded
to $128$ under encryption---an exact, no-loss operation (not rank reduction), so FHE and plaintext
scores are identical (correlation $1.0$). We do \emph{not} hold out a separate tuning set: the
rank and beam-width operating points, and the face decision threshold, are chosen on these
same queries, so they are best-case operating points---the recall and latency
\emph{trends} we rely on are smooth across the sweeps, but the single reported points
should be read as such.

\subsection{Face corpus (222k centroids from 10\,M images, 512-dim)}
\label{sec:eval:face}
\begin{figure}[t]
  \centering
  \includegraphics[width=0.49\linewidth]{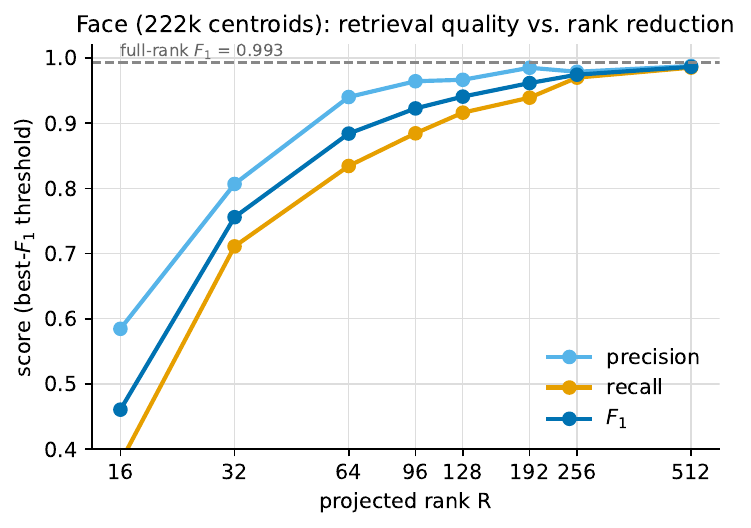}\hfill
  \includegraphics[width=0.49\linewidth]{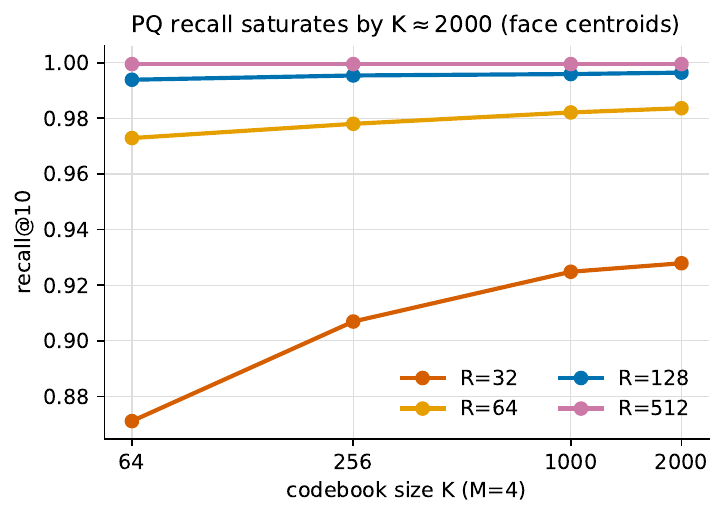}
  \caption{Face. \emph{Left:} quality vs.\ rank (Table~\ref{tab:face}).
  \emph{Right:} PQ recall@10 saturates by $K\!\approx\!2000$; very low rank ($R{=}32$)
  caps quality regardless of $K$.}
  \label{fig:face}
\end{figure}

This corpus uses \emph{no hierarchy}: at $222$k centroids we score the encrypted query
against \emph{all} of them (a flat membership scan), so there is no coarse/leaf routing
and no access-pattern leakage---the hierarchy of \S\ref{sec:method:hier} is used only for the
billion-scale corpora (DataComp and Deep1B). Retrieval is judged against query \emph{identity}: we cluster
${\sim}10$\,M face images (Glint360K) down to
${\sim}\num{222049}$ identity centroids and report precision/recall/$F_1$ of a
threshold on cosine (no reranking---a cluster either matches or not). This is the
setting that most stresses rank reduction (512-dim). Table~\ref{tab:face} and
Fig.~\ref{fig:face} show quality degrading gracefully with rank: rank~256 keeps
$F_1{=}0.974$ against the full-rank $0.993$, and $F_1$ only falls off a cliff below
rank~64. PQ on top saturates by codebook size $K\!\approx\!2000$.

\begin{table}[t]
  \centering
  \caption{Face (222k centroids): precision/recall/$F_1$ at the best-$F_1$ threshold,
  and warm per-query FHE latency (steady-state, one-time enrollment excluded), vs.\
  projected rank. Latency was measured at
  $R\in\{32,64,128\}$ and grows steeply with rank.}
  \label{tab:face}
  \begin{tabular}{lrrrr}
    \toprule
    rank & precision & recall & $F_1$ & latency/q \\
    \midrule
    full (512) & 0.991 & 0.995 & \textbf{0.993} & --- \\
    512 & 0.988 & 0.985 & 0.986 & --- \\
    256 & 0.979 & 0.970 & 0.974 & --- \\
    192 & 0.985 & 0.939 & 0.962 & --- \\
    128 & 0.967 & 0.916 & 0.941 & 7.0\,s \\
    96  & 0.964 & 0.884 & 0.923 & --- \\
    64  & 0.940 & 0.834 & 0.884 & 2.1\,s \\
    32  & 0.807 & 0.711 & 0.756 & 0.8\,s \\
    16  & 0.585 & 0.380 & 0.461 & --- \\
    \bottomrule
  \end{tabular}
\end{table}

\subsection{DataComp-1B ($1.39\times10^9$, 512-dim CLIP)}
\begin{figure}[t]
  \centering
  \includegraphics[width=0.49\linewidth]{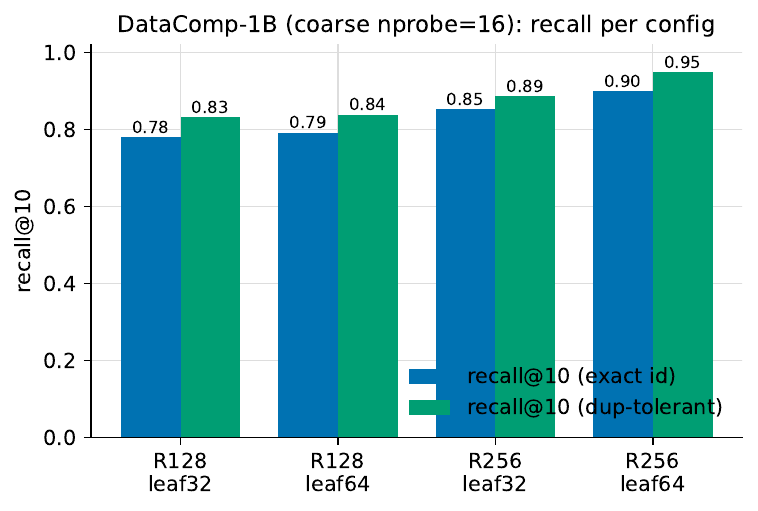}\hfill
  \includegraphics[width=0.49\linewidth]{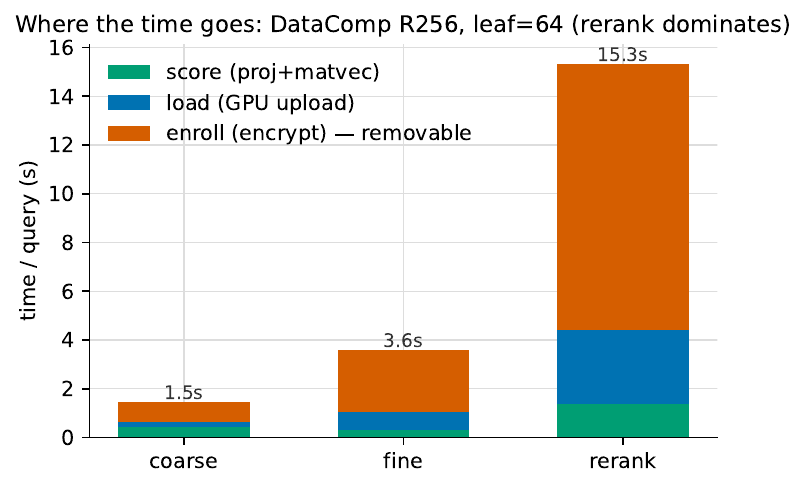}
  \caption{DataComp-1B. \emph{Left:} recall@10 (exact and duplicate-tolerant) per
  configuration. \emph{Right:} where the per-query (cold) time goes for R256/leaf-$64$---the
  rerank stage dominates, and per-query enrollment (encrypting the candidate rows) is the
  removable gap to the deployable warm number (load $+$ score).}
  \label{fig:datacomp}
\end{figure}

This is the headline setting: full-scale, image-to-image, using the hierarchy with
rank reduction to $256/128$. Because the data is web-scraped and full of near-
duplicates, we report both \emph{exact-id} recall@10 (is the labeled answer in our
top-10) and \emph{duplicate-tolerant} recall@10 (does our top-10 contain a
$\ge0.95$-cosine near-duplicate of it)---the latter is the honest quality on this
data. In fact, at full rank the duplicate-tolerant recall equals \emph{containment}
(was the answer's leaf loaded at all): every exact-id miss is a $\ge0.99$-cosine
near-duplicate, so the reranking is effectively perfect and the remaining gap is
purely routing---this is the $0.975$ ceiling in the plaintext reference row of
Table~\ref{tab:datacomp}. Table~\ref{tab:datacomp} and Fig.~\ref{fig:datacomp} give
every configuration:
rank~256 reaches \textbf{0.90}/\textbf{0.95} (exact/dup), rank~128 a lighter
\textbf{0.79}/\textbf{0.84}, and the coarse beam saturates at $16$.

\begin{table}[t]
  \centering
  \caption{DataComp-1B: recall@10 and \emph{warm} per-query latency per configuration
  (coarse beam $16$). Warm latency $=$ loading the pre-encrypted candidate ciphertexts and
  scoring them (disk, $16$ threads, plaintext pre-rotation); the database is encrypted once,
  offline. It excludes the prototype's on-the-fly per-query enrollment (encryption), which
  roughly triples the time (the \emph{cold} cost). The top row is
  the plaintext, no-rank-reduction reference---the accuracy ceiling (its exact-id misses are
  all near-duplicates, hence dup-tol $=$ containment).}
  \label{tab:datacomp}
  \begin{tabular}{llrrr}
    \toprule
    rank & leaf & recall@10 exact & recall@10 dup-tol & warm latency/q \\
    \midrule
    512 (plaintext ref) & 64 & 0.910 & 0.975 & N/A \\
    \midrule
    128 & 32 & 0.780 & 0.833 & 1.8\,s \\
    128 & 64 & 0.792 & 0.839 & 2.7\,s \\
    256 & 32 & 0.853 & 0.887 & 4.0\,s \\
    256 & 64 & \textbf{0.900} & \textbf{0.950} & \textbf{6.1}\,s \\
    \bottomrule
  \end{tabular}
\end{table}

\subsection{Deep1B ($10^{9}$, 96-dim)}
\label{sec:eval:deep1b}
Deep1B uses the \emph{same} coarse\,$\to$\,fine\,$\to$\,rerank hierarchy as DataComp, scored under
encryption at every level with the client picking from decrypted scores---but at a much lower
\emph{native} dimension. The $96$-dim vectors are zero-padded to $128$ and scored directly with
\emph{no} rank reduction and \emph{no} projection: the pad is an exact identity, so the FHE score
equals the plaintext $96$-dim cosine (correlation $1.0$). Table~\ref{tab:deep1b} sweeps the coarse and
leaf beams; the operating point (coarse $16$, leaf $64$) reaches recall@10 $0.906$, with high-recall
points to $0.974$. Running the \emph{entire} pipeline under all-levels FHE on $2000$ queries reproduces
this: recall@10 $\mathbf{0.9045}$ at a \emph{warm} per-query latency of $\mathbf{2.30}$\,s (load
$1.64$\,s $+$ score $0.66$\,s; the cold prototype, with on-the-fly enrollment, is $7.6$\,s), scoring
$1000$ coarse $+$ ${\sim}18$k fine $+$ ${\sim}75$k rerank candidates. DataComp needs its R256/leaf-$64$
configuration ($6.1$\,s warm; Table~\ref{tab:datacomp}) to reach the same $0.90$ recall, so Deep1B
gets there ${\sim}2.6\times$ faster---chiefly because the native $128$-dim packing is denser than
DataComp's zero-padded $512$-dim path (Deep1B is also a lower-dimensional, and thus easier, corpus).

\begin{table}[t]
  \centering
  \caption{Deep1B ($10^{9}$, 96-dim): recall@10 (plaintext routing) vs.\ coarse and leaf beam, $2000$
  queries. The $96{\to}128$ zero-pad is exact, so all-levels FHE reproduces this---the operating point
  (coarse $16$, leaf $64$, \textbf{bold}) gives FHE recall@10 $0.9045$ at $2.3$\,s warm per query.}
  \label{tab:deep1b}
  \begin{tabular}{rrrrrr}
    \toprule
     & \multicolumn{5}{c}{leaf beam} \\
    coarse beam & $16$ & $32$ & $64$ & $128$ & $256$ \\
    \midrule
    $4$  & 0.706 & 0.767 & 0.797 & 0.814 & 0.819 \\
    $8$  & 0.762 & 0.829 & 0.873 & 0.896 & 0.909 \\
    $16$ & 0.781 & 0.856 & \textbf{0.906} & 0.936 & 0.953 \\
    $32$ & 0.790 & 0.870 & 0.921 & 0.955 & 0.974 \\
    \bottomrule
  \end{tabular}
\end{table}

\subsection{Latency and scaling}
\begin{figure}[t]
  \centering
  \includegraphics[width=0.49\linewidth]{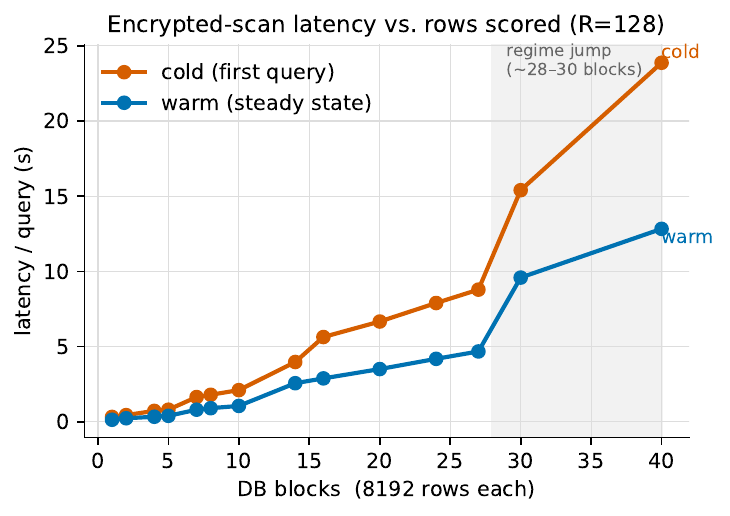}\hfill
  \includegraphics[width=0.49\linewidth]{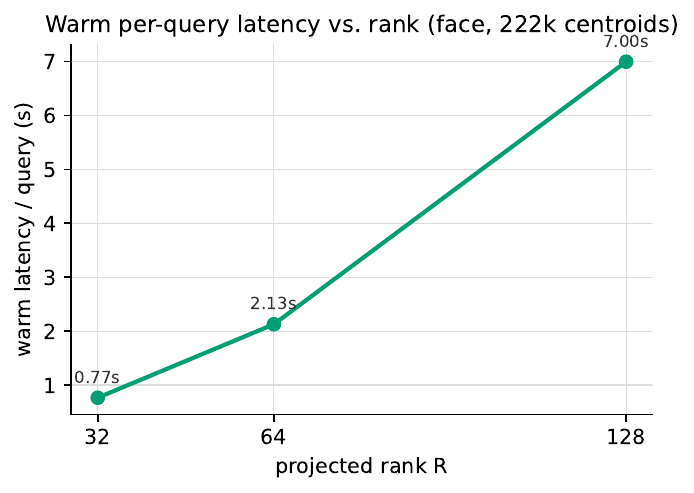}
  \caption{\emph{Left:} encrypted-scan latency (cold and warm) vs.\ number of scored rows,
  from a standalone R128 scan (thread settings independent of Table~\ref{tab:datacomp}).
  \emph{Right:} warm per-query latency vs.\ rank (face, 222k).}
  \label{fig:scaling}
\end{figure}

Fig.~\ref{fig:scaling} shows how encrypted-scan latency grows with the number of rows
scored (super-linearly, with a jump once the working set outgrows a GPU threshold),
and the warm per-query cost as a function of rank. Together with the optimization
stack (Fig.~\ref{fig:ablation}) these set the few-seconds-per-query operating point.

% =====================================================================
\section{Access-Pattern Leakage}
\label{sec:leakage}
Because the server computes only on ciphertexts, it never sees the query, the scores,
or the returned answer. What it does see is the \emph{access pattern}: which group ids
the client asks it to gather at each level. We describe what that reveals, level by
level, in our coarse\,$\to$\,leaf hierarchy with no oblivious-access protection.

Throughout we assume a \emph{strong} attacker: because a single search is several
round trips (a coarse request, then a leaf request that depends on the coarse
result), the attacker must be able to tell that these requests belong to the same
query and link them into one session---by timing, ordering, or a session
identifier---and to link a query to its predecessors. A weaker attacker that only
saw an interleaved, unlinkable stream of gather requests from many concurrent queries
could not attribute a leaf request to the coarse selection that produced it, and would
learn correspondingly less. The analysis below is therefore an upper bound on what a
realistic server-side observer learns.

\paragraph{What the server stores (and therefore what it can observe).} We store each
level of the index as an independent flat file of all embeddings at that level; no tree
and no child\,$\to$\,parent map are stored anywhere on the server. The coarse\,$\to$\,leaf
linkage exists only on the client, which computes it after decrypting each level's
scores. As a result the client reveals a \emph{selection} to the server twice per query.
First, after scoring the coarse level (a full scan of the coarse file, which leaks
nothing on its own), it fetches the leaves \emph{under} its chosen coarse cells so they
can be scored---the server sees that set of leaf ids, the query's coarse neighborhood.
Second, after scoring those leaves, it fetches the database rows in the few leaves it
keeps---the server sees the fine leaf selection. Both are just sets of leaf ids in the
flat leaf file; the server never sees a coarse id or which coarse cell a leaf belongs to.
This flat, tree-free storage is the sole structural assumption behind the analysis below,
and it is what keeps Parent Recovery (below) a quantity the attacker must \emph{recover},
rather than one storage hands it for free.

\paragraph{What the attacker sees, level by level.}
\begin{itemize}
  \item \textbf{Coarse scoring.} The client fetches the entire coarse file on every query
        (all ${\sim}1000$ cells are scored), so this step reveals nothing about which
        cells the query preferred.
  \item \textbf{Coarse pick (first reveal).} To score the leaf level, the client fetches
        the leaves that are \emph{children} of the coarse cells it chose. The server sees
        that set of leaf ids---the query's coarse neighborhood---but, with flat storage
        and no parent map, not which coarse cell they came from. It can only
        \emph{recover} the coarse\,$\to$\,leaf grouping from co-access across many queries
        (Parent Recovery, below).
  \item \textbf{Leaf pick (second reveal).} The client then fetches the database rows in
        the few leaves it keeps; the server sees those leaf ids---the query's fine
        localization.
\end{itemize}
So each query reveals two nested locations: a coarse neighborhood (the leaves under its
coarse pick) and, inside it, the fine leaf set it keeps. It never reveals the query
vector, the scores, the coarse ids themselves, or which reranked row is the answer.

\subsection{Quantifying the leak}
The interesting leak is not only \emph{where is this one query} but \emph{how the
queries relate to each other}: two access sets that overlap tell the server the two
private queries live in the same part of the space, even though it never learns either
vector. This ``relationships between access sets'' view is the classic lesson of the
searchable-encryption literature~\cite{islam2012access}, and a recent vector-search
benchmark makes the same point for IVF search---revealing which clusters a query probes
is a privacy property in its own right, separate from hiding the vector and the
scores~\cite{vecsearchbench2026}. Everything below is measured \emph{offline in
plaintext} from a held-out query set. The router gives each query $q$ two access sets:
its \emph{coarse pick} $A(q)$ (the leaves under its chosen coarse cells) and its
\emph{leaf pick} $S(q)$ (the few leaves it keeps, $|S(q)|=b$, the leaf beam). The
query-side metrics below use the sharper leaf pick $S(q)$; the database-side metrics use
both. We use the plaintext query vectors only to score against ground truth: the query
similarity $\mathrm{sim}(q_i,q_j)=\langle \hat q_i,\hat q_j\rangle$ (cosine of the
normalized queries) and the true nearest-neighbor and parent structure. No attacker is
trained and no ciphertext is touched.

\paragraph{Access-pattern defenses.} We compare three ways the client can pad the sets
it fetches, applied at \emph{both} reveals---the coarse pick and the leaf pick, each with
its own group size $G$ (write $O$ for a padded set, standing for either $A(q)$ or
$S(q)$). With \emph{no padding} the server observes the true set directly. With
\emph{random padding} it observes $O=S\cup D$, where the decoy set $D$ is drawn fresh on
every query. With \emph{seeded padding} every cell $r$ has a fixed anonymity group $g(r)$
of size $G$, and the client always fetches $O=\bigcup_{r\in S} g(r)$---the same set on
every repeat. We hold the observed size equal across the two schemes so the comparison is
not confounded by one simply fetching more, and we sweep the coarse and leaf group sizes
$(G_{\mathrm{coarse}}, G_{\mathrm{leaf}})$ independently to see how each reveal's padding
moves each metric.

\paragraph{The intersection attack.} Padding helps only if the decoys cannot be told
apart from the real cells. Against the strong attacker assumed above---one that can link
$t$ repeats of the same query---the server intersects the observations,
\[
  R_t(q) = \bigcap_{i=1}^{t} O_i(q),
\]
which removes anything that changes between repeats. Fresh random decoys appear in only
one observation and wash out, so $R_t(q)\!\to\!S(q)$; a fixed seeded group is present
every time and survives, so $R_t(q)=\bigcup_{r\in S(q)} g(r)$ for all $t$. This one
mechanism drives every comparison below: random padding is undone by repetition, seeded
padding is not (it instead lowers the resolution to the group). Where queries never
repeat or cannot be linked, the intersection collapses to $t=1$ and the two schemes are
indistinguishable.

\paragraph{Query-side metrics.} For a query $q$ (or its cleaned set $R_t(q)$ under
repeats):
\begin{itemize}
  \item \textbf{Absolute localization} $|S(q)|=b$: the nearest leaf is one of $b$
        candidates. The single-query baseline.
  \item \textbf{Relative localization}
        $J(S(q_i),S(q_j))=\dfrac{|S(q_i)\cap S(q_j)|}{|S(q_i)\cup S(q_j)|}$ plotted
        against $\mathrm{sim}(q_i,q_j)$: does access overlap track query similarity? A
        strong relationship means the access pattern exposes the \emph{geometry} of the
        private query stream---the most telling of the three, since similarity is exactly
        what an embedding is meant to encode. We report the leaf-set Jaccard binned by
        query cosine, and its rank correlation.
  \item \textbf{Repeated-observation narrowing}
        $R_t(q)=\big|\bigcap_{i=1}^t O_i(q)\big|$ vs.\ $t$: how many candidates survive
        repeated observation. Fresh padding decays toward $b$; seeded padding stays flat
        at the group size.
  \item \textbf{Access-pattern uniqueness}
        $U(q)=\big|\{q'\neq q : J(S(q'),S(q))\ge \tau\}\big|$: an anonymity-set
        size---how many other queries share, to within overlap $\tau$, the same trace
        ($\tau=1$ is exact equality).
\end{itemize}

\paragraph{Database-side metrics.} Over a stream the server also accumulates co-access.
With $n_i=\#\{q: i\in O(q)\}$ and $c_{ij}=\#\{q: i,j\in O(q)\}$, the normalized
co-access is $J_{ij}=c_{ij}/\sqrt{n_i n_j}$ (computed on the cleaned sets $R_t$ when
repeats are available).
\begin{itemize}
  \item \textbf{Neighbor Recovery@}$k$: rank each cell's neighbors by $J_{ij}$ and
        compare to its true $k$ nearest cells,
        \[
          \mathrm{NR@}k=\frac{1}{N}\sum_i
            \frac{|\,\mathrm{top}^{J}_k(i)\cap \mathrm{NN}^{\mathrm{true}}_k(i)\,|}{k},
        \]
        computed on the leaf picks $S(q)$ (a leaf neighbor graph) and on the coarse picks
        $A(q)$ (a coarse neighbor graph). Only the coarse \emph{scoring} is a full scan
        with no signal; the coarse \emph{pick} does leak, so both levels are recoverable.
        Because the attacker is never handed the tree, there is no static, query-free
        version---recovery exists only through the observed co-access.
  \item \textbf{Parent Recovery}: cluster the co-access of the coarse picks $A(q)$---leaves
        fetched together because they share a coarse parent---and score the clusters
        against the true leaf\,$\to$\,coarse grouping (adjusted Rand index). Because we
        store flat files with no parent map, this grouping is never given to the
        server---it is a genuine recovered quantity here, not the trivial $1.0$ it would
        be in a tree-storing deployment.
\end{itemize}

\paragraph{What the defenses buy.} Under the intersection attack the two schemes
separate cleanly. Random padding reverts every metric to its no-padding value as $t$
grows: the overlap--similarity correlation reappears, $R_t$ decays to $b$, uniqueness
returns, and Neighbor and Parent Recovery climb back to baseline. Seeded padding holds
against repetition, but only by \emph{lowering the resolution}---overlaps collapse to
group granularity, $R_t$ plateaus at the group size, uniqueness rises to the group's
occupancy, and recovery flattens at a floor set by the group construction (most
effective when the groups are dispersed, for Neighbor Recovery, or straddle true
parents, for Parent Recovery). It never removes the leak; it quantizes it. This is the
same conclusion the searchable-encryption literature reaches for repeated and linkable
searches~\cite{oya2021hiding,osse2021}: hiding the access pattern one query at a time is
not enough once the attacker can link queries over time.
\subsection{Measured leakage}
\label{sec:leak-measured}
We evaluate all of the above offline on the $10{,}000$ held-out DataComp image queries at
the operating point (coarse beam $16$, leaf beam $64$, $10^{6}$ leaves). Query
similarity is the cosine of the plaintext query embeddings; the geometry ground truth is
the leaf and coarse centroids. Rank is fixed---the leak is a property of the routing, not
of the projection.

\paragraph{How each quantity is computed, and its approximations.} Everything is a
function of the observed access sets and the plaintext query vectors; no attacker is
trained. \emph{Co-access} between two cells is $J_{ij}=c_{ij}/\sqrt{n_i n_j}$, where
$n_i$ is the number of queries that fetch cell $i$ and $c_{ij}$ the number that fetch
both (the cosine of their query-membership vectors). \emph{Neighbor Recovery@}$k$
($k{=}10$) ranks each cell's partners by $J_{ij}$ and reports the mean overlap of that
top-$k$ with the cell's true $k$ nearest centroids by cosine. \emph{Parent Recovery}
clusters the co-access graph by linking each leaf to its single strongest partner and
taking connected components, scored against the true leaf\,$\to$\,coarse map by adjusted
Rand index. \emph{Precision} samples query pairs and reports the mean query cosine
conditioned on sharing ${\ge}d$ cells, with lift relative to the random-pair mean.
\emph{Repeated-observation narrowing} re-draws the padding on a query's fixed routing
$t$ times and intersects the observations. Three approximations keep this tractable and
are unbiased or clearly bounded. \emph{(i)}~The co-access \emph{graph} is exact (every
co-occurring pair is counted), but the Neighbor-Recovery \emph{average} is estimated over
a sample of ${\sim}4000$ well-connected leaves rather than all $10^{6}$---a Monte-Carlo
estimate of a mean, not a subsample of the graph (a uniform sample of $10^{6}$ leaves
almost never co-occurs, giving an empty graph); the coarse side uses all ${\sim}1000$
coarse cells, and Parent Recovery is scored over the ${\sim}3000$ most-accessed leaves.
\emph{(ii)}~For large padded sets we cap the co-access pair count by using fewer queries;
since $J_{ij}$ is scale-invariant under query subsampling (numerator and both
denominators scale together), this changes the estimator's variance, not its value.
\emph{(iii)}~The similarity metrics use ${\sim}0.1$--$0.5$M sampled query pairs rather
than all $\binom{10^{4}}{2}$; a shared leaf fires for only ${\sim}0.1\%$ of pairs, so
those conditional means rest on a few hundred pairs.

\paragraph{Reading the numbers.} Neighbor Recovery and Parent Recovery are fractions in
$[0,1]$: $1$ means the attacker reconstructs the true neighbor graph (or the true
leaf\,$\to$\,coarse grouping) \emph{exactly} from access patterns alone, and $0$ means no
better than chance---for Neighbor Recovery@$10$ over $n$ cells chance is ${\approx}10/n$,
i.e.\ ${\sim}10^{-5}$ over $10^{6}$ leaves and ${\sim}0.01$ over $1000$ coarse cells, so a
Coarse Neighbor Recovery of $0.01$ under padding is effectively dead. The similarity
\emph{lift} is the mean query cosine given a shared cell divided by the mean over random
pairs: lift $=1$ means a shared cell says nothing about similarity, and lift $>1$ means
the two queries are more alike than a random pair by that factor---so a shared leaf at
lift $1.57$ makes a pair $57\%$ more similar than chance, and seeded padding driving the
lift toward $1$ is the signal being destroyed. Finally $R_t$ is how many candidate cells
survive $t$ repeated observations: the size of the attacker's residual hiding set.

\begin{figure}[t]\centering
\includegraphics[width=0.49\linewidth]{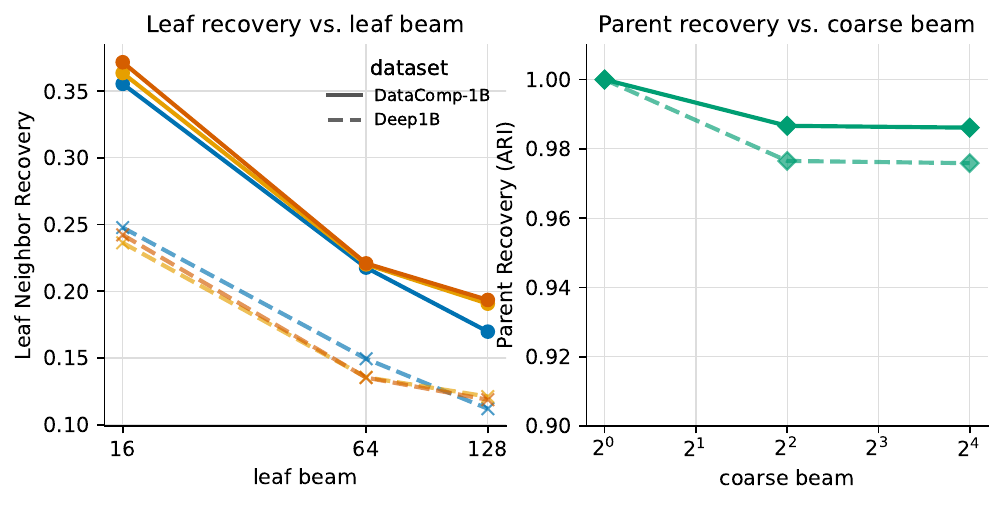}\hfill
\includegraphics[width=0.49\linewidth]{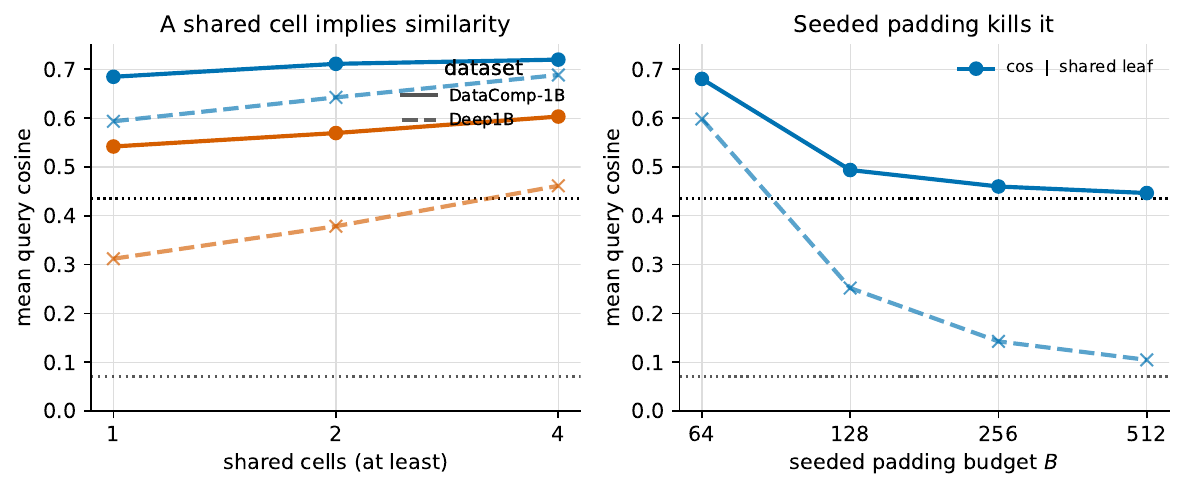}
\caption{\emph{Left:} leakage vs.\ beam width---Leaf Neighbor Recovery falls as the leaf
beam widens (coarse-independent), Parent Recovery is near-perfect. \emph{Right:} a shared
leaf implies higher query similarity than a shared coarse cell, and seeded padding drives
that cosine to the random-pair baseline. Solid: DataComp-1B; dashed: Deep1B---whose
random-pair cosine is much lower ($0.07$ vs.\ $0.44$), so the same shared-cell signal is a
far larger \emph{lift} ($8.4\times$ vs.\ $1.6\times$ for a shared leaf).}
\label{fig:leak-precision}
\end{figure}

\paragraph{What leaks with no defense (Table~\ref{tab:leak-base}).} The database geometry
leaks substantially from co-access alone: the attacker recovers $72\%$ of each coarse
cell's true nearest neighbors, and reconstructs the leaf\,$\to$\,coarse grouping almost
perfectly (adjusted Rand index $0.99$)---because a query fetches a coarse cell's whole
leaf block, same-parent leaves are inseparable in the access log. Leaf-level neighbor
recovery is lower ($0.22$) but non-trivial. The stream also leaks \emph{query} similarity:
a shared coarse cell raises the expected pairwise cosine from $0.44$ (a random pair) to
$0.54$, and a shared \emph{leaf}---far rarer---to $0.68$ (Fig.~\ref{fig:leak-precision}).
A shared leaf is thus the sharper signal: rare but high-confidence.
On \textbf{Deep1B} the picture holds structurally---Coarse Recovery $0.75$, Parent $0.98$, Leaf $0.14$,
all within a few points of DataComp (Table~\ref{tab:leak-base})---but the query-similarity \emph{lift}
is far larger: a shared leaf makes a pair $8.4\times$ more similar than random (vs.\ $1.6\times$ on
DataComp) and a shared coarse cell $4.4\times$, because Deep1B's random-pair cosine ($0.07$) is much
lower, so the same shared-cell signal is proportionally stronger. The recovery numbers are thus a
property of the routing, not the corpus; the lift depends on how spread out the query distribution is.

\begin{table}[t]\centering\small
\begin{tabular}{lrr}
\toprule
quantity & DataComp-1B & Deep1B \\
\midrule
Leaf Neighbor Recovery & $0.22$ & $0.14$ \\
Coarse Neighbor Recovery & $0.72$ & $0.75$ \\
Parent Recovery (adjusted Rand index) & $0.99$ & $0.98$ \\
mean cosine, random query pair & $0.44$ & $0.07$ \\
mean cosine $\mid$ shared leaf (lift) & $0.68\ (1.57\times)$ & $0.59\ (8.4\times)$ \\
mean cosine $\mid$ shared coarse cell (lift) & $0.54\ (1.24\times)$ & $0.31\ (4.4\times)$ \\
\bottomrule
\end{tabular}
\caption{Baseline access-pattern leakage, no defense, for both billion-scale corpora (coarse beam
$16$, leaf beam $64$, $10^{6}$ leaves, $10{,}000$ queries; routing scored at full embedding
dimension---$512$ for DataComp, $96$ for Deep1B---so the leak is a property of the plaintext
geometry, independent of the FHE rank reduction $R$). Recovery is nearly identical across
datasets---the leak is a property of the routing, not the data---but the query-similarity \emph{lift}
is far larger on Deep1B, whose random-pair cosine ($0.07$) is much lower than DataComp's ($0.44$), so a
shared cell is a proportionally stronger signal.}
\label{tab:leak-base}
\end{table}

\begin{figure}[t]\centering
\includegraphics[width=0.49\linewidth]{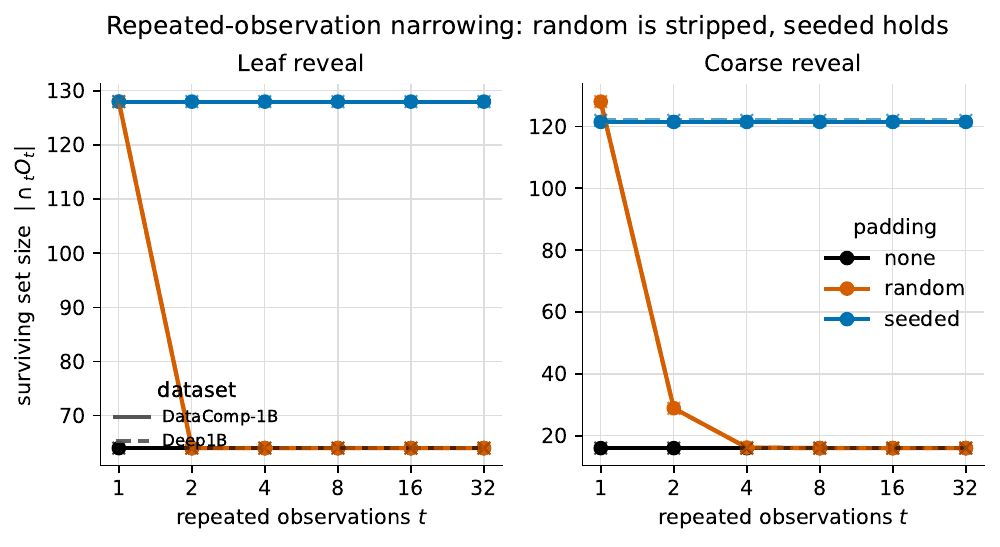}\hfill
\includegraphics[width=0.49\linewidth]{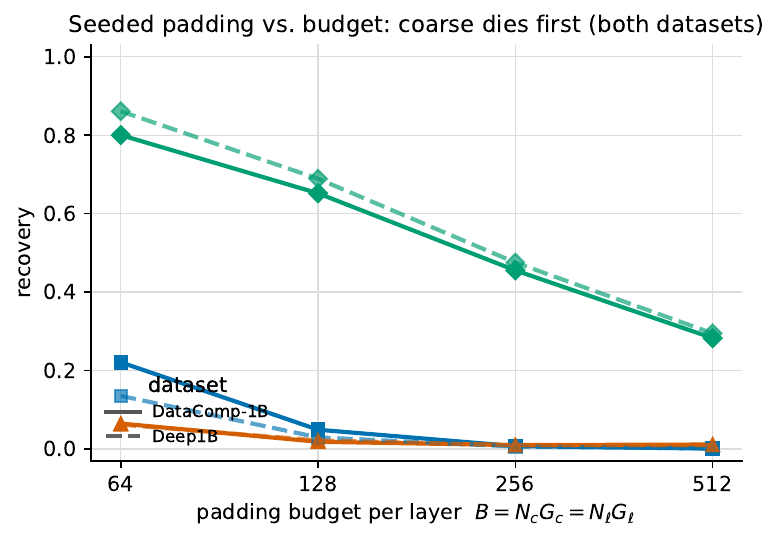}
\caption{\emph{Left:} repeated-observation narrowing---random decoys are stripped by the
intersection while seeded groups survive, at both reveals. \emph{Right:} recovery vs.\
padding budget on both reveals; the coarse leaks die first. Solid: DataComp-1B; dashed: Deep1B
(the two nearly coincide---the leak is a property of the routing, not the corpus).}
\label{fig:leak-defense}
\end{figure}

\paragraph{Random vs.\ seeded padding under the intersection attack
(Table~\ref{tab:leak-defense}, Fig.~\ref{fig:leak-defense}).} We pad both
reveals---the coarse pick and the leaf pick---with fresh (random) or fixed-group (seeded)
decoys, and let the attacker intersect repeated queries. Random padding hides every leak
at a single observation, but the intersection strips it and Leaf Neighbor Recovery, Coarse
Neighbor Recovery, and the cosine signal all revert to baseline. Seeded padding holds
under repetition---Coarse Neighbor Recovery collapses from $0.72$ to $0.02$, Parent
Recovery from $0.99$ to $0.65$, and stays there. (Random padding never dents Parent
Recovery, even at a single observation: fresh decoys are noise the clustering averages
out, whereas the fixed seeded groups corrupt it.)

\begin{table}[t]\centering\small
\begin{tabular}{llrrrrr}
\toprule
scheme & attack & Leaf Rec. & Coarse Rec. & Parent Rec. & lift$\mid$leaf & lift$\mid$coarse \\
 & & \multicolumn{5}{c}{\footnotesize DataComp-1B\,/\,Deep1B} \\
\midrule
random & no intersection & $0.08$/$0.04$ & $0.52$/$0.45$ & $0.98$/$0.98$ & $1.1$/$2.1$ & $1.0$/$1.0$ \\
random & with intersection & $0.22$/$0.14$ & $0.72$/$0.75$ & $0.99$/$0.98$ & $1.6$/$8.5$ & $1.2$/$4.4$ \\
seeded & no intersection & $0.05$/$0.03$ & $0.02$/$0.02$ & $0.65$/$0.69$ & $1.2$/$3.4$ & $1.0$/$1.1$ \\
seeded & with intersection & $0.05$/$0.03$ & $0.02$/$0.02$ & $0.65$/$0.69$ & $1.1$/$3.3$ & $1.0$/$1.1$ \\
\bottomrule
\end{tabular}
\caption{Random vs.\ seeded padding on both reveals, with and without the intersection
attack ($B{=}128$), reported as \textbf{DataComp-1B\,/\,Deep1B}. Random is undone by repetition
(reverts to baseline); seeded holds. lift$\mid$leaf and lift$\mid$coarse are the query-similarity
lifts (mean cosine of a pair sharing that cell, over the random-pair mean); seeded padding drives both
toward $1$. The coarse lift is weak on DataComp ($\le\!1.2\times$) but sizeable on Deep1B ($4.4\times$
at baseline); the leaf lift is the sharp signal ($8.5\times$ on Deep1B), which seeded padding cuts to
$3.3\times$.}
\label{tab:leak-defense}
\end{table}

\begin{figure}[t]\centering
\includegraphics[width=0.62\linewidth]{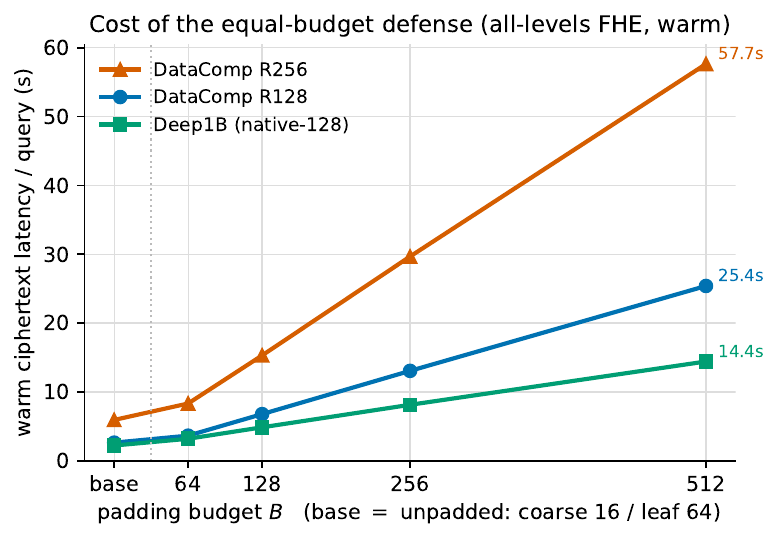}
\caption{Warm ciphertext latency per query vs.\ padding budget $B$ for the three configurations
(all-levels FHE, load$+$score). \emph{base} is the unpadded operating point (coarse $16$, leaf
$64$); each budget $B$ fetches $B$ coarse cells and $B$ leaves, so latency grows
\emph{approximately linearly} in $B$ (the fetch cost, not exponential)---the same dial that
collapses the leak in Table~\ref{tab:leak-budget}. Axis is linear in $B$.}
\label{fig:budget-latency}
\end{figure}

\paragraph{Equal-budget defense (Table~\ref{tab:leak-budget},
Fig.~\ref{fig:leak-defense}).} Under a shared padding budget $B=N_c G_c=N_\ell G_\ell$ (base
$N_c{=}16$, $N_\ell{=}64$, so $G_c{=}B/16$ and $G_\ell{=}B/64$) the
coarse side is far cheaper to hide: since $N_c\!\ll\!N_\ell$, the same budget buys
$G_c=4G_\ell$, so the coarse leak dies quickly (Coarse Neighbor Recovery
${\approx}0.02$ by $B{=}128$) while the leaf leak needs much more before $G_\ell$
even exceeds $1$. As $B$ grows, Parent Recovery falls $0.80\!\to\!0.28$, Leaf Neighbor
Recovery $0.22\!\to\!0.001$, and the shared-leaf cosine $0.68\!\to\!0.45$ (the signal is
dead). Seeded padding never removes a leak; it \emph{quantizes} it to group granularity at
a fetch-cost multiplier---a clean privacy/cost dial, cheapest on the coarse side. Its cost in
\emph{warm ciphertext latency}---the full padded pipeline scored end-to-end under FHE---is
Table~\ref{tab:budget-latency} and Fig.~\ref{fig:budget-latency}.

\begin{table}[t]\centering\small
\begin{tabular}{rrrrrrrr}
\toprule
$B$ & $G_c$ & $G_\ell$ & Leaf Rec. & Coarse Rec. & Parent Rec. & lift$\mid$leaf & lift$\mid$coarse \\
 & & & \multicolumn{5}{c}{\footnotesize DataComp-1B\,/\,Deep1B} \\
\midrule
$64$ & $4$ & $1$ & $0.22$/$0.14$ & $0.06$/$0.06$ & $0.80$/$0.86$ & $1.6$/$8.4$ & $1.0$/$1.3$ \\
$128$ & $8$ & $2$ & $0.05$/$0.03$ & $0.02$/$0.02$ & $0.65$/$0.69$ & $1.1$/$3.5$ & $1.0$/$1.1$ \\
$256$ & $16$ & $4$ & $0.007$/$0.006$ & $0.01$/$0.01$ & $0.46$/$0.48$ & $1.0$/$2.0$ & $1.0$/$1.0$ \\
$512$ & $32$ & $8$ & $0.001$/$0.001$ & $0.01$/$0.01$ & $0.28$/$0.30$ & $1.0$/$1.5$ & $1.0$/$1.0$ \\
\bottomrule
\end{tabular}
\caption{Equal-budget seeded padding on both reveals (with intersection), as
\textbf{DataComp-1B\,/\,Deep1B}. Coarse is cheapest to hide because $G_c=4G_\ell$. As the budget grows
both query-similarity lifts collapse toward $1$ (the signal is destroyed): the leaf lift falls
$1.6\!\to\!1.0$ on DataComp and $8.4\!\to\!1.5$ on Deep1B, while the (already weak) coarse lift is
${\approx}1$ throughout on DataComp and $1.3\!\to\!1.0$ on Deep1B. Raw cosines are in
Table~\ref{tab:leak-base} and Fig.~\ref{fig:leak-precision}.}
\label{tab:leak-budget}
\end{table}

\begin{table}[t]\centering\small
\setlength{\tabcolsep}{4.5pt}
\begin{tabular}{llccrrrrr}
\toprule
 & coarse/ & recall & dup- & \multicolumn{5}{c}{Warm ciphertext latency (s) at budget $B$} \\
\cmidrule(lr){5-9}
Configuration & leaf & @10 & tol & base & $64$ & $128$ & $256$ & $512$ \\
 & & & & {\scriptsize$(1,1)$} & {\scriptsize$(4,1)$} & {\scriptsize$(8,2)$} & {\scriptsize$(16,4)$} & {\scriptsize$(32,8)$} \\
\midrule
DataComp R128 & $16/64$ & $0.792$ & $0.839$ & $2.63$ & $3.64$ & $6.77$ & $13.06$ & $25.39$ \\
DataComp R256 & $16/64$ & $0.900$ & $0.950$ & $5.93$ & $8.31$ & $15.30$ & $29.67$ & $57.65$ \\
Deep1B ($128$) & $16/64$ & $0.905$ & \textit{--} & $2.21$ & $3.19$ & $4.86$ & $8.13$ & $14.42$ \\
\bottomrule
\end{tabular}
\caption{\textbf{Cost of the equal-budget defense} (the dial in Table~\ref{tab:leak-budget},
plotted in Fig.~\ref{fig:budget-latency}). Warm (load$+$score, database pre-encrypted---the
deployable cost) per-query latency of the \emph{full padded} pipeline with \emph{all} routing
levels under FHE (coarse pick, leaf pick, and rerank all encrypted). \emph{base} is the unpadded
operating point (coarse $N_c{=}16$, leaf $N_\ell{=}64$); at budget $B$ the padded fetch is exactly
$B$ coarse cells and $B$ leaves (route \texttt{level0:}$B$,\,\texttt{fine:}$B$), with copies
$(G_c,G_\ell){=}(B/16,\,B/64)$ shown under each column. recall@10 and dup-tolerant recall@10 are at
the base operating point---padding fetches decoys and does not change what is retrieved. Base
latencies agree with Table~\ref{tab:datacomp} ($2.7$/$6.1$\,s) and the Deep1B $2.30$\,s operating
point to within run-to-run variance (e.g.\ $5.93$ vs.\ $6.1$\,s at R256); every measurement passed
the plaintext golden check (correlation $1.0$). Deep1B uses exact-NN ground
truth (no near-duplicate notion, hence no dup-tol). One L40S.}
\label{tab:budget-latency}
\end{table}

% =====================================================================
\section{Related Work}
\label{sec:related}
We build on the CKKS approximate-arithmetic scheme (OpenFHE) accelerated on GPU by
FIDESlib, whose baby-step/giant-step diagonal linear transforms our projection and
scoring extend to the rank-reduced, hierarchical setting. Closest in spirit is recent
GPU-accelerated encrypted face-similarity search with a rotation-key-efficient
BSGS-Diagonal packing, under the same single-key model, evaluated at up to
${\sim}32$K templates~\cite{demicheli2026encface}; earlier encrypted-matching systems
(HERS, Blind-Match, GROTE) target biometric one-to-many matching. We differ by
targeting billion-scale ANN and by combining rank reduction and a hierarchy to
get there.

% =====================================================================
\section{Conclusion and Future Work}
\label{sec:conclusion}
We showed that encrypted-query ANN scales to a billion vectors on a GPU by composing
rank reduction and a hierarchical index, reaching recall@10 of
$0.90$/$0.95$ (exact/duplicate-tolerant) on DataComp-1B at ${\sim}6$\,s per encrypted
(warm) query (server-side), and recall@10 $0.90$ on the $96$-dim Deep1B benchmark
(all-levels FHE, $2.3$\,s warm per query), and we documented the full protocol and every
configuration's cost.
We also measured the system's \textbf{access-pattern leakage}
(\S\ref{sec:leakage}): even though the values are encrypted, the hierarchy's access
pattern reveals which cells are fetched, and from co-access alone an observer recovers
much of the database geometry (coarse-neighbor recovery $0.72$ on DataComp-1B and $0.75$
on Deep1B, parent-grouping recovery $0.98$--$0.99$---a property of the routing, not the
data) and a weaker query-similarity signal. Fixed-group (``seeded'') padding suppresses
these durably where fresh-decoy padding is undone by a repeated-query intersection
attack---at a fetch-cost that is cheapest on the coarse reveal. Tightening this
privacy/cost trade-off, and closing the gap to a fully oblivious traversal, is the main
direction we leave open.
On the performance side, the \emph{warm} cost is dominated by loading candidate ciphertexts
onto the GPU (Table~\ref{tab:budget-latency}, Fig.~\ref{fig:datacomp}), and the access
distribution is heavily skewed---a small set of clusters is routed to far more often than the
rest. A straightforward win is therefore to \emph{minimize loads by pinning the hot clusters
resident on-GPU}: keep the most-frequently-fetched clusters permanently encrypted and loaded on
dedicated GPUs (sharding the index across devices by fetch frequency), so common queries skip the
reload entirely. The fetch-frequency profile needed to do this is already available from the
access log---trivially so, since it is the very same co-access signal we study for leakage---which
makes this a low-effort systems optimization, orthogonal to the cryptography.
In the same vein, since the dominant load cost is paid per fetched region rather than per query,
\emph{batching many queries} against each loaded region---scoring a whole batch in one pass before
evicting it---would amortize that load over the batch and raise throughput (slot packing already
amortizes \emph{within} a query; this extends the idea \emph{across} queries).
Complementary on the accuracy side, the per-query cost scales with the working dimension (the
reduced rank $R$ for DataComp, native $128$ for Deep1B), so \emph{stronger compression}---learned
or quantized projections, product or residual quantization---that preserves recall at an even lower
dimension would cut the dominant scoring cost directly; how far the dimension can be pushed before
recall degrades is an open question worth pursuing.
Finally, the encrypted database's \emph{storage and upload} footprint---not just its per-query load
cost---can be shrunk with orthogonal cryptographic tools: switching each stored ciphertext down to the
minimal modulus its remaining computation needs~\cite{brakerski2012leveled}, regenerating the
public-random half of a fresh ciphertext from a short seed, and \emph{transciphering} (hybrid
homomorphic encryption)---uploading data under a compact symmetric cipher and homomorphically
converting it to FHE server-side, which removes the ciphertext-expansion cost from the
wire~\cite{naehrig2011can,canteaut2016stream,cho2021transciphering}---each trading a one-time
server-side computation for a large reduction in stored and transmitted ciphertext.
The deepest lever is the modulus chain itself: a ciphertext's size is set by its number of RNS
limbs, which is bounded below by the multiplicative depth the score requires (depth $4$ here). Once
each stored row already sits at that minimal level, modulus switching buys nothing further---so the
remaining win is to \emph{lower the depth requirement itself}, via a shallower scoring circuit, which
would shrink every stored ciphertext's limb count and hence both its storage and its dominant load
cost across the entire database.

% =====================================================================
\bibliographystyle{plain}
\bibliography{references}

\end{document}